# The concept of weak measurements and the super-efficiency of quantum tomography

Yu.I. Bogdanov[a,b,c], N.A. Bogdanova[a,b], B. I. Bantysh[a,b], Yu. A. Kuznetsov[a,b]

a. Valiev Institute of Physics and Technology, Russian Academy of Sciences, Russia
b. National Research University of Electronic Technology (MIET), Russia
c. National Research Nuclear University (MEPhI), Russia

**ABSTRACT**

The quantum measurement procedure based on the Lorentz transformation formalism and weak perturbation of the system is considered. In the simple case of a single-qubit it turns out that one can perform 4-dimension pseudo-rotation along with ordinary 3-dimension rotations on the Bloch sphere. These pseudo-rotations are similar to the Lorentz transformation in special relativity theory. The extension of the Lorentz transformation for many-qubit systems is also considered. The quantum measurement protocols based on the Lorentz transformation are proposed. It has been shown that these protocols cease to form the decomposition of unity and could be superefficient providing the fidelity higher than any POVM-measurement protocol. However, one can perform the complement of the Lorentz protocol to POVM-protocol by an additional measurement operator. If the initial mixed state is close to the pure one this operator corresponds to weak perturbation of the state while the original Lorentz protocol sets the strong perturbations. As the result, the feedback provides an effective control of a quantum system introducing weak perturbations to the quantum state.

The results of this research are essential for the development of methods for the control of quantum information technologies.

## 1. INTRODUCTION

The spinor representation of the Lorentz group is a fundamental tool for theoretical and mathematical physics. It is based on the fact that the complex matrix *L* of dimension $2 \times 2$ and unit determinant generates the transformation that maintains the relativistic interval, i.e. the Lorentz transformation[1,2]. Here we consider the Lorentz formalism as an application for the description of quantum transformations and measurements of qubits and systems of a higher dimension. This paper is the extension of our results reported at [3,4].

Let us consider the Lorentz group spinor representation for the mixed state of a qubit. As this state can be considered as a mixture of two components, one can define the corresponding purified state $\varphi_{in}$ which has a form of a complex matrix of dimension $2 \times 2$ (each column corresponds to one of the mixture components). In this case the initial density matrix is

$$\rho_{in} = \varphi_{in} \cdot \varphi_{in}^{+} \qquad (1)$$

Let us note that the purification procedure is ambiguous. This property plays a fundamental role for quantum tomography and measurements[5-7]. In general, one can consider the qubit density matrix as a mixture of $n \geq 2$ components so $\varphi_{in}$ has the dimension $2 \times n$. One can also multiply the matrix $\varphi_{in}$ to the right by an arbitrary unitary matrix $U$ of dimension $n \times n$ ($\varphi_{in} \to \varphi_{in} U$) as this transformation keeps the density matrix (1) unchanged. Note that the matrix of amplitudes $\varphi_{in}$ could be stretched into a column of length $2n$. However, for our purposes, it is more convenient to represent the purified state $\varphi_{in}$ in the form of a matrix of dimension $2 \times n$.

The spinor transformation is defined by a complex matrix $L$ of dimension $2 \times 2$ with the unit determinant:

$$\varphi_{out} = L\varphi_{in}, \det(L) = 1 \qquad (2)$$

The output density matrix has the form

$$\rho_{out} = \varphi_{out} \cdot \varphi_{out}^+ = L \cdot \rho_{in} \cdot L^+ \qquad (3)$$

The density matrix determinant remains unchanged under this transformation. This fact provides the invariance of the relativistic interval. Indeed, one can decompose any density matrix of a qubit by the identity matrix $I$ of dimension $2 \times 2$ and three Pauli matrices $\sigma_1, \sigma_2, \sigma_3$ [8]:

$$\rho = \frac{1}{2}\left(P_0 \cdot I + \vec{P}\vec{\sigma}\right) \qquad (4)$$

In terms of quantum optics, $P_0 = Tr(\rho)$ corresponds to the radiation intensity and $P_1 = Tr(\rho\sigma_1)$, $P_2 = Tr(\rho\sigma_2)$, $P_3 = Tr(\rho\sigma_3)$ are Stokes parameters. Note that here we consider the density matrix normalized to the radiation intensity but not to unity.

For the standard representation of Pauli matrices

$$\rho = \frac{1}{2}\begin{pmatrix} P_0 + P_3 & P_1 - iP_2 \\ P_1 + iP_2 & P_0 - P_3 \end{pmatrix} \qquad (5)$$

One can consider the four-vector $(P_0, \vec{P}) = (P_0, P_1, P_2, P_3)$ as the Minkowski-Stokes vector. The invariance of the density matrix determinant under the spinor transformation results in the invariance of the value $s^2 = P_0^2 - \vec{P}^2$, which plays the role of the squared relativistic interval. It is easy to see, that the determinant of the density matrix is proportional to the mass $m$ of the effective particle under consideration:

$$\det(\rho) = \frac{1}{4}\left(P_0^2 - \vec{P}^2\right) = \frac{1}{4}m^2 \qquad (6)$$

Note, that this spinor representation of the Lorentz group has found its application in optics for many different uses[9-13]. Let us consider a real Lorentz transformation matrix $G$ of dimension $4 \times 4$ for the Minkowski-Stokes vector: $P'_\mu = G_{\mu\nu} P_\nu$, $\mu, \nu = 0, 1, 2, 3$. In a detailed view, this transformation could be expressed in the form

$$\begin{pmatrix} P'_0 \\ P'_1 \\ P'_2 \\ P'_3 \end{pmatrix} = \frac{1}{2} \begin{pmatrix} \alpha^*\alpha + \beta^*\beta + \gamma^*\gamma + \delta^*\delta & \alpha^*\beta + \beta^*\alpha + \gamma^*\delta + \delta^*\gamma & i(\alpha^*\beta - \beta^*\alpha + \gamma^*\delta - \delta^*\gamma) & \alpha^*\alpha - \beta^*\beta + \gamma^*\gamma - \delta^*\delta \\ \alpha^*\gamma + \beta^*\delta + \gamma^*\alpha + \delta^*\beta & \alpha^*\delta + \beta^*\gamma + \gamma^*\beta + \delta^*\alpha & i(\alpha^*\delta - \beta^*\gamma + \gamma^*\beta - \delta^*\alpha) & \alpha^*\gamma - \beta^*\delta + \gamma^*\alpha - \delta^*\beta \\ -i(\alpha^*\gamma + \beta^*\delta - \gamma^*\alpha - \delta^*\beta) & -i(\alpha^*\delta + \beta^*\gamma - \gamma^*\beta - \delta^*\alpha) & \alpha^*\delta - \beta^*\gamma - \gamma^*\beta + \delta^*\alpha & -i(\alpha^*\gamma - \beta^*\delta - \gamma^*\alpha + \delta^*\beta) \\ \alpha^*\alpha + \beta^*\beta - \gamma^*\gamma - \delta^*\delta & \alpha^*\beta + \beta^*\alpha - \gamma^*\delta - \delta^*\gamma & i(\alpha^*\beta - \beta^*\alpha - \gamma^*\delta + \delta^*\gamma) & \alpha^*\alpha - \beta^*\beta - \gamma^*\gamma + \delta^*\delta \end{pmatrix} \begin{pmatrix} P_0 \\ P_1 \\ P_2 \\ P_3 \end{pmatrix} \quad (7)$$

which depends on the complex matrix elements of $L$:

$$L = \begin{pmatrix} \alpha & \beta \\ \gamma & \delta \end{pmatrix} \qquad (8)$$

Matrix $G$ underlines the Lorentz tensor transformations, which, as one could see above, has a more complex form than the spinor representation of Lorentz group. One could suggest, that the spinors are more fundamental objects than four-vectors in a coordinate or momentum space (e.g. such ideas form the basis of Penrose twistor methods[14,15]). Yet, for our purpose it is important that it is applicable to qubits and, in particular, to the polarization degree of freedom of the electromagnetic field. The generally multicomponent spinor representation of polarization $\varphi$ is obviously more fundamental in comparison to the Minkowski-Stokes vector $(P_0, \vec{P})$.

Let us consider the simple Lorentz transformation

$$L = \begin{pmatrix} \exp(-\eta/2) & 0 \\ 0 & \exp(\eta/2) \end{pmatrix} \qquad (9)$$

with the real parameter $\eta$. In this case, the new Minkowski-Stokes components depend on the initial values as

$$\begin{aligned} P'_0 &= \cosh(\eta) P_0 - \sinh(\eta) P_3, \\ P'_3 &= -\sinh(\eta) P_0 + \cosh(\eta) P_3, \\ P'_1 &= P_1, \quad P'_2 = P_2 \end{aligned} \qquad (10)$$

This transformation describes the transition to the new frame of reference that moves along the $z$ axis with the velocity $v = \tanh(\eta)$ relatively to the original one (we consider the speed of light equal to unity). Note that

$$\cosh(\eta) = \frac{1}{\sqrt{1-v^2}}, \quad \sinh(\eta) = \frac{v}{\sqrt{1-v^2}} \qquad (11)$$

One can interpret the considered transformation as the pseudo-rotation in the plane $(t, z)$. One can similarly consider the pseudo-rotation in $(t, x)$ and $(t, y)$ planes. Clearly, the Lorentz group also contains regular rotations in $(x, y), (x, z)$ and $(y, z)$ planes.

## 2. QUANTUM MEASUREMENTS FOR SINGLE-QUBIT STATES

The velocity of an effective particle with Minkowski-Stokes vector $(P_0, \vec{P})$ is $\vec{v} = \vec{P}/P_0$, where $P_0$ is the energy of this effective particle and $\vec{P}$ is its momentum. Let us consider the transformation that performs the transition to the frame of relevance with zero momentum (the center of mass frame). In this case the qubit is located in the center of the Bloch sphere. Such a transformation is provided by the following pseudo-rotation operator:

$$L_{\vec{n}}(\theta) = \exp\left(-\frac{\theta}{2}\vec{\sigma}\vec{n}\right) = \cosh\left(\frac{\theta}{2}\right) I - \sinh\left(\frac{\theta}{2}\right)\vec{\sigma}\vec{n} \tag{12}$$

Here $v = \tanh(\theta)$ is the absolute value of the effective particle velocity and $\vec{n}$ is its direction. This equation describes the Lorentz boost – relative motion of two systems of reference with a constant velocity and without the rotation of their coordinate systems. Operator (12) provides the generalization of the transformation that we have discussed in the previous section. Note, that one could obtain such a transformation for mixed states only. This transition is impossible for pure qubit states as well as one cannot proceed to the frame of reference of a photon using relativistic Lorentz transformation.

Let us consider the procedure of the Minkowski-Stokes vector measurement. Let the laser beam propagate through a polarizer with the vertical transmission axis. The number of photons $N_V$ that have passed the polarizer within the fixed time interval (we assume this interval to be equal to unity) are being registered. This measurement corresponds to the measurement of the projective operator (projector) $|V\rangle\langle V|$. After the rotation of the polarizer by $90^0$ degrees one can also measure the number of photons $N_V$ which corresponds to the orthogonal projector $|H\rangle\langle H|$. The total number of photons $N_V + N_H$ serves as the estimation of the component $P_0 = Tr(\rho)$ of the Minkowski-Stokes vector, while the difference of photon numbers $N_V - N_H$ gives the estimation of the component $P_3 = Tr(\rho\sigma_3)$. Let the photons undergo a unitary rotation $U$ before passing through the polarizer. Then the third component of the Stokes vector is

$$P_3' = Tr(U\rho U^+ \sigma_3) = Tr(\rho U^+ \sigma_3 U) = Tr(\rho \sigma_3') \tag{13}$$

where $\sigma_3' = U^+ \sigma_3 U$. For the vertical polarizer transmission axis the transformation $U$ results in the projection of the input state on $U^+|V\rangle$. This situation corresponds to the measurement of the projector $U^+|V\rangle\langle V|U$. Similarly, if the polarizer transmission axis is horizontal one obtains the measurement of the projector $U^+|H\rangle\langle H|U$.

Obviously, to measure $P_3$, one should perform the identity transformation, i.e. $U = I$. Operators

$$U_1 = \frac{1}{\sqrt{2}}\begin{pmatrix} 1 & 1 \\ 1 & -1 \end{pmatrix}, \quad U_2 = \frac{1}{\sqrt{2}}\begin{pmatrix} 1 & -i \\ 1 & i \end{pmatrix} \tag{14}$$

provide the projective measurements of observables $\sigma_1$ and $\sigma_2$ respectively and, as a result, the estimation of $P_1$ and $P_2$ components of the Stokes vector. Here and below we assume that

$$|V\rangle = \begin{pmatrix} 1 \\ 0 \end{pmatrix}, \quad |H\rangle = \begin{pmatrix} 0 \\ 1 \end{pmatrix} \tag{15}$$

Note, that these transformation provide the corresponding transformation of Pauli matrices as $\sigma_1 = U_1^+ \sigma_3 U_1$ and $\sigma_2 = U_2^+ \sigma_3 U_2$. It is easy to perform operations $U_1$ and $U_2$ by optical phase plates.

Let us generalize the above procedure by performing the Lorentz boost transformation of the initial state. It is especially important to consider the transition to the center of mass frame of a qubit that is initially in an arbitrary mixed state. It turns out that in this frame, which corresponds to the qubit located in the center of the Bloch sphere, the fidelity of the quantum state tomography is the highest (for the fixed sample size)[7,16]. Let us introduce the instrumental matrix $X$ to describe the quantum tomography protocol. Each row $X_j$ of this matrix defines the bra-vector in the Hilbert space, $j = 1, …, m$, where $m$ is the number of the protocol rows (total number of measured projections).

The probability amplitude for a state vector $c$ and $j$-th protocol row is $M_j = X_j c$. The square of its absolute value determines the expected number of registered events

$$\lambda_j = |M_j|^2 = \langle c|\Lambda_j|c\rangle \tag{16}$$

where $\Lambda_j = X_j^+ X_j$ is the measurement operator. For the density matrix $\rho$ one would have

$$\lambda_j = \mathrm{Tr}(\Lambda_j \rho) \tag{17}$$

The measurement protocol contains each row of the instrumental matrix with the weight $t_j$ (exposure time). We will normalize the joint weighted probability to the overall sample size $n$:

$$\sum_{j=1}^{m} t_j \lambda_j = n \tag{18}$$

If the measurement protocol forms the decomposition of unity, one would have

$$\sum_{j=1}^{m} t_j \Lambda_j = nI \tag{19}$$

Note, that equation (18) is clear from (19) for the state vectors, normalized to unity: $\langle c|c\rangle = 1$. To consider the Lorentz transformation of the input quantum state one can transform the initial instrumental matrix $X^{in}$ as follows:

$$X^{out} = X^{in} L \tag{20}$$

We assume that all rows of the initial instrumental matrix are normalized to unity: $X_j^{in}(X_j^{in})^+ = 1$, $j = 1, 2, ..., m$. However, after the non-unitary transformation $L$ this normalization is changed. We can interpret the weights $t_j = X_j^{out}(X_j^{out})^+$ as exposure times for new measurement protocol if we assume the initial exposure times equal to unity.

The accuracy of the reconstruction of the quantum state $\rho_0$ is determined by fidelity[17,18]:

$$F = \left(Tr\sqrt{\sqrt{\rho}\rho_0\sqrt{\rho}}\right)^2 \tag{21}$$

where $\rho$ is the reconstructed density matrix. It can be shown[7,19] that for any POVM-measurements the average fidelity losses $\langle 1-F \rangle$ cannot be lower than the following value:

$$\langle 1-F \rangle_{min} = \frac{f^2}{4n(s-1)} \tag{22}$$

where $f = (2s - r)r - 1$ and $r$ are the number of degrees of freedom and rank of the density matrix and $s$ is the Hilbert space dimension.

One can consider the efficiency of a quantum measurement protocol as the ratio between $\langle 1 - F \rangle_{min}$ and the average fidelity losses one could obtain for a given quantum state and this protocol:

$$eff = \frac{\langle 1 - F \rangle_{min}}{\langle 1 - F \rangle} \tag{23}$$

Generally speaking, as the Lorentz protocol does not form the decomposition of unity, it could violate the condition (22) on the minimum average fidelity losses. It turns out that this type of measurements could result in the emergence of superefficiency with $eff > 1$.

Let us consider the mixed quantum state, that is almost pure. Let the principal component of the state density matrix have the weight $\lambda_0 = 0.99$ and be equal to

$$c_0 = \begin{pmatrix} \cos(\theta/2)\exp(-i\varphi/2) \\ \sin(\theta/2)\exp(i\varphi/2) \end{pmatrix}, \quad \theta = \pi/4, \quad \varphi = 5\pi/3 \tag{24}$$

The corresponding density matrix is

$$\rho = \begin{pmatrix} 0.84648 & 0.17324 + 0.30006i \\ 0.17324 - 0.30006i & 0.15351 \end{pmatrix} \tag{25}$$

In this case $P_0 = 1$ and the velocity of the effective particle coincides with the state Bloch vector and is close to the speed of light:

$$v = \lambda_0 - \lambda_1 = 2\lambda_0 - 1 = 0.98, \quad \lambda_0 + \lambda_1 = 1 \tag{26}$$

The initial instrumental matrix, based on the tetrahedron symmetry, and exposure times are

$$X = \begin{pmatrix} 0.82047 + 0.33985i & 0.42471 - 0.17592i \\ -0.42471 + 0.17592i & -0.82047 - 0.33985i \\ 0.17592 + 0.42471i & 0.33985 - 0.82047i \\ -0.33985 + 0.82047i & -0.17592 - 0.42471i \end{pmatrix} \begin{matrix} t_1 = 1 \\ t_2 = 1 \\ t_3 = 1 \\ t_4 = 1 \end{matrix} \tag{27}$$

After the Lorentz transformation to the center of mass frame the instrumental matrix and exposure times are

$$X = \begin{pmatrix} 0.27927 + 0.36463i & 0.54059 - 0.70486i \\ 0.19104 - 0.19964i & -0.90837 - 0.31388i \\ 0.2146 + 0.32545i & 0.38736 - 0.83545i \\ 0.13795 + 0.43802i & 0.26592 - 0.84758i \end{pmatrix} \begin{matrix} t_1 = 3.7506 \\ t_2 = 4.2893 \\ t_3 = 9.7821 \\ t_4 = 2.2788 \end{matrix} \tag{28}$$

It can be shown that the considered Lorentz transformation strongly deforms the measurement protocol so the projectors are located in a tiny sector of the Bloch sphere opposite to the location of the state $\rho$.

Figure 1 depicts the agreement between the results of the tomography numerical simulation with Lorentz protocol and the theoretical distribution, which is based on the results developed in our previous works[5-7,16,19].

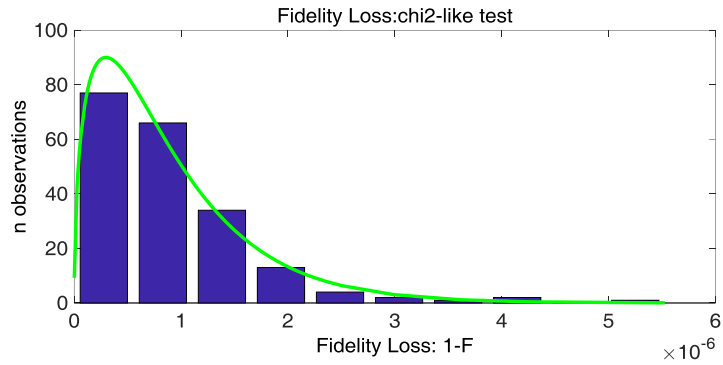

Figure 1. Distribution of fidelity losses. Comparison of the theory (curve) and numerical experiments (histogram) for the state (17) and Lorentz measurement protocol (28). The number of numerical experiments performed is 200. In each experiment 100 thousand state representatives have been measured.

The considered protocol turns out to be superefficient. The efficiency (or superefficiency) (23) is:

$$\text{superefficiency} = \gamma^2 = 1/(1-f^2) = 25.25(25) \tag{29}$$

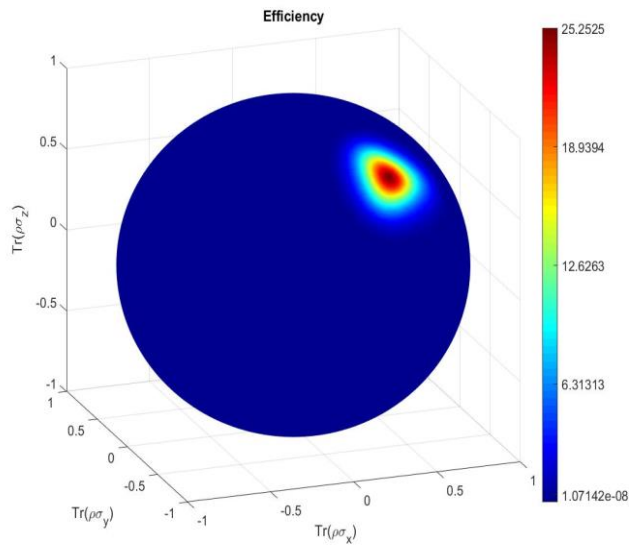

Figure 2. The efficiency distribution on the Bloch sphere for the Lorentz protocol (28). The maximum and minimum efficiency are $\text{eff}_{\max} = 25.25(25)$ and $\text{eff}_{\min} \approx 1.87 \cdot 10^{-5}$

Figure 2 demonstrates the local nature of the superefficiency. The considered Lorentz protocol provides extremely high fidelity only for quantum states within a quite small region nearby $\rho$. For all the other states on the Bloch sphere the fidelity turns out to be very low. The minimum average fidelity loss is more than million times lower than the maximum one. In order to tune the protocol for a desired state one should use the adaptive procedure: the measurements start with any standard protocol followed by gradual transition to the Lorentz protocol during the collection of statistical data. We call this phenomenon superefficiency, by analogy with superefficiency in classical statistics[20,21]. The superefficient statistical estimator achieves lower asymptotic variance than any regular efficient estimator. However, superefficient estimator could be better than a regular efficient estimator on a set of measure zero only.

It is remarkable, that the total exposure time is increased by the factor of $\gamma = 1/\sqrt{1-f^2}$ for the Lorentz protocol (28) and the similar ones (similarly to the relativistic time dilation). However, let us note, that in the case of the Lorentz protocol parameters $t_j$ in (28) describe weights of the instrumental matrix rows rather than the exposure time itself. These weights describe the relative intensity with which measurement operators influence the measured state. This kind of measurements for polarizing quantum states could be carried out with use of a partially-polarizing beam splitter[22-24].

In fact, the Lorentz protocol describes some kind of unfinished measurements. This means that passing the photon through a set of partially-polarizing beam splitters, corresponding to the Lorentz protocol, does not necessarily lead to its registration. Let us consider this fact in detail and complete the measurement protocol $X_L = X^{out}$ (see eq. (20)) to make it form the decomposition of unity with the following operator:

$$A_0 = a_0 I - X_L^+ X_L \tag{30}$$

where $a_0$ is the maximal eigenvalue of the matrix $X_L^+ X_L$. By definition, $A_0$ is positive and $A_0 + X_L^+ X_L = a_0 I$ so the resulting protocol forms the POVM-measurements.

Note, that $X_L^+ X_L = \sum_{j=1}^m t_j \Lambda_j$ where $\Lambda_j$ is the projector corresponding to the j-th row of the instrumental matrix with $m$ rows, and $\Lambda_j^2 = \Lambda_j$.

Let us denote

$$\Lambda_0 = \frac{A_0}{a_0} = I - \frac{1}{a_0} \sum_{j=1}^m t_j \Lambda_j \tag{31}$$

Then the operator $\Lambda_0$, together with the operators, $\frac{t_j}{a_0} \Lambda_j$, $j = 1, 2, ..., m$, form the decomposition of unity. In this case

$$p_j = Tr\left(\rho \frac{t_j}{a_0} \Lambda_j\right), j = 1, 2, \ldots, m \tag{32}$$

is the probability that the photon will be absorbed by the j-th detector during the measurement of the initial state $\rho$. Similarly, $p_0 = Tr(\rho \Lambda_0)$ is the probability that the photon will not be absorbed by any of the detectors of the Lorentz protocol. Moreover, since the considered set of $m+1$ operators forms the POVM-protocol

$$p_0 + \sum_{j=1}^{m} p_j = 1 \tag{33}$$

Let us now return to the state (17) with $\lambda_0 = 0.99$, $\lambda_1 = 0.01$, v=0.98 and perform the above procedure of Lorentz protocol complement for the tetrahedron ($m = 4$) and cube ($m = 6$) protocols[6,7]. Below we give the fidelity between the initial and reduced quantum states for each of the m+1 rows of the new protocol. It turns out that measurement operators $\frac{t_j}{a_0} \Lambda_j$, $j=1,2,\ldots,m$ describe strong perturbations of the input state (only 2% of the representatives are affected).

Operator $\Lambda_0$ corresponds to the 5th and 7th rows and describes 98% of all measurements and performs weak perturbation of the input state. It is remarkable that the reduced state resulting from the action of the measurement operator corresponds to the initial pure state.

$$Tetrahedron = \begin{bmatrix} 0.026529 \\ 0.023197 \\ 0.010172 \\ 0.043663 \\ 0.99 \end{bmatrix}, \quad Cube = \begin{bmatrix} 0.064488 \\ 0.011695 \\ 0.030298 \\ 0.014705 \\ 0.012374 \\ 0.049515 \\ 0.99 \end{bmatrix}$$

Thus, the presented results show that the additional measurement operator corresponds to a weak perturbation of the initial state, while all the measurement operators of the original Lorentz protocol perform strong perturbations. The protocol turns out to be superefficient if we use only strongly perturbed representatives of the quantum statistical ensemble in the total sample size. We stress that only these strongly perturbed representatives are registered while all the weakly perturbed states leave the system undetected.

It is important to note that Lorentz measurements provide both a precise analyzer and an optimal polarizer of a quantum state. As an analyzer, the Lorentz protocol performs the precise measurement of the initial mixed state that is close to a pure one. As an optimal polarizer, the Lorentz protocol gives a pure state with the largest possible number of representatives (this state corresponds to the main component of the original density matrix).

## 3. LORETNZ QUANTUM MEASUREMENTS FOR STATES OF AN ARBITARY DIMENSION

We have considered the Lorentz transformation for a single qubit, but this approach can be generalized for the case of quantum systems of an arbitrary higher dimension. In this case, the instrumental matrix transformation to the center of mass frame is also carried out by (20): $X^{out} = X^{in} L$.

The Lorentz transformation matrix is constructed according to the following equation:

$$L = \frac{1}{\sqrt{s}} \psi_{in}^{-1}, \quad L \to \frac{L}{(\det L)^{1/s}} \tag{34}$$

Here $s$ is the dimension of the Hilbert space, $\psi_{in}$ - is the matrix of amplitudes for purified state $\rho_{in}$ ($\rho_{in} = \psi_{in}\psi_{in}^+$). Matrix $\psi_{in}$ has the dimension $s \times s$. Note that the second relation in (35) provides the condition $\det L = 1$.

Matrix *L* of the form (35) defines the generalized Lorentz transformation for the state in *s*-dimensional Hilbert space. This matrix acts on states in an *s* - dimensional Hilbert space. We do not associate this transformation with any tensor transformation in the four-dimensional space-time.

It can be shown that the density matrix after the transformation is $\rho_{out} = (\det \rho_{in})^{1/s} I$, where *I* is the identity matrix of dimension $s \times s$.

The example shown on the Figure 3 corresponds to the tomography of the two-qubit state. The original measurement protocol specifies a set of mutually unbiased bases[25]. Again, Figure 3 illustrates the agreement between the results of the tomography numerical simulation with Lorentz protocol and the theoretical distribution

For the simulation, we have used the state with the following weights: $\lambda_1 = 0.99$, $\lambda_2 = \lambda_3 = \lambda_4 = 1/300$. The superefficiency of the protocol is $47.032$.

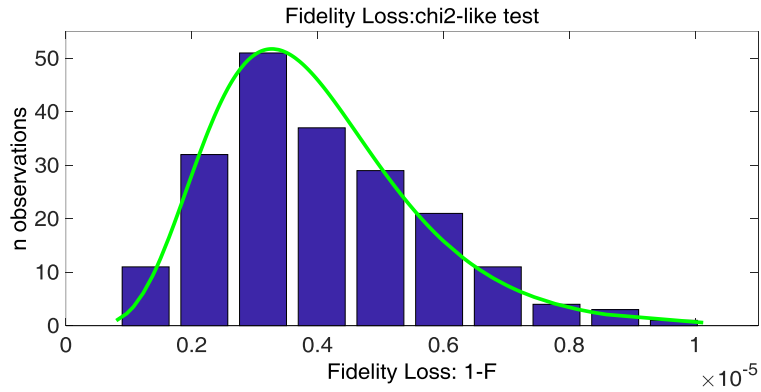

Figure 3. Distribution of fidelity losses for two-qubit state. Comparison of the theory (curve) and numerical experiments (histogram) for the Lorentz measurement protocol.

Note, that the above description is suitable for systems of an arbitrarily high dimension. In this case, the Lorentz transformation matrix is constructed according to the general rule (30). Any protocol with the property of tomographic completeness[6,7,16] can be chosen as the initial measurement protocol.

# CONCLUSION

Let us briefly formulate the main results of this work.

It has been shown that along with three-dimensional rotations of a quantum state on the Bloch sphere, there exist four-dimensional pseudo-rotations, which are similar to the Lorentz transformations in special relativity theory.

The proposed quantum measurement protocols based on Lorentz transformations do not form the decomposition of unity. Such protocols can be superefficient, that is, they can provide higher accuracy than any standard POVM-measurement protocol.

Lorentz protocols describe unfinished measurements. In particular, this means that passing the photon through a set of partially-polarizing beam splitters, corresponding to the Lorentz protocol, does not necessarily leads to its registration.

One could complement any Lorentz protocol to a POVM-measurement protocol by introducing an additional measurement operator. This operator corresponds to a weak perturbation of the initial mixed state that is close to a pure one while the measurement operators of the original Lorentz protocol correspond to strong perturbations. If one considers only these strongly perturbed states in the total sample size, the measurement protocol turns out to be superefficient. These states are being registered by detectors while others leave the system unregistered and could be used for further performing of quantum algorithms.

One can consider the Lorentz protocols both as a precision analyzer and an optimal polarizer of the optical quantum state. As an analyzer, Lorentz measurements perform the precise tomography of the initial mixed state that is close to a pure one. As an optimal polarizer, the Lorentz protocol provides for a pure state with the largest possible number of representatives.

Along with single-qubit systems, the Lorentz transformation theory could be applied to systems with an arbitrarily high dimension.

The results of the study are essential for the development of optimal methods for controlling quantum states and operations.

# ACKNOWLEDGEMENT

This work was supported by Russian Foundation of Basic Research (project 18-37-00204).